# Simulating the Dynamics of T Cell Subsets Throughout the Lifetime

Stephanie Foan, Andrew Jackson, Ian Spendlove, Uwe Aickelin

Academic Unit of Clinical Oncology and Intelligent Modelling and Analysis Research Group,
University of Nottingham

**Abstract.** It is widely accepted that the immune system undergoes age-related changes correlating with increased disease in the elderly. T cell subsets have been implicated. The aim of this work is firstly to implement and validate a simulation of T regulatory cell ($T_{reg}$) dynamics throughout the lifetime, based on a model by Baltcheva. We show that our initial simulation produces an inversion between precursor and mature $T_{rey}$s at around 20 years of age, though the output differs significantly from the original laboratory dataset. Secondly, this report discusses development of the model to incorporate new data from a cross-sectional study of healthy blood donors addressing balance between $T_{rey}$s and Th17 cells with novel markers for $T_{reg}$. The potential for simulation to add insight into immune aging is discussed.

## 1 Introduction to System Dynamics Modelling of Immunity

Simulation has been defined as methods and applications mimicking the behaviour of a real system [1]. The benefits of simulation to immunology include time- and cost-effectiveness as well as less labour- and resource-intensiveness resulting from removal from the biological environment. *In vitro* experimentation is useful for investigating individual interactions but is far removed from the whole picture, and *in vivo* experimentation is useful for the whole picture but is unlikely to answer specific questions [2]. Using simulation, flexibility is available for systematically generating hypotheses and conducting experiments impossible to do practically, yet informed by robust data and literature.

System dynamics simulations are useful for looking at complex systems over time. They are characterised by stocks of an entity and flows between stocks [3]. Immune system examples of stocks include precursor and mature T cell pools and flows might represent transition of cells from precursor to mature. This technique is useful for modelling relationships defined by differential equations. An example of a differential equation describing T cell dynamics is the change in number of precursor cells equated to proliferation of precursors minus death and maturation rates. This ongoing work will apply system dynamics simulation technique to complement *in vitro* studies of Th 17s and $T_{reg}$ s throughout the lifetime.

## 2 The Need for Balance: Th 17s and $T_{rey}$s Throughout the Lifetime

The immune system maintains a balance between mounting an adequate immune response to protect from infection and restricting the size of the immune response to prevent damage to self. There is evidence to suggest an age-related tendency to a pro-inflammatory environment [4] contributing to more collateral damage and autoimmune diseases. This work addresses the hypothesis that important contributors to this state of imbalance are Th17 cells (amplifying immune responses) and $T_{reg}$s (dampening down immune responses). Although some studies have shown an age-related increase in the number of $T_{reg}$s in human peripheral blood [5][6], it has also been shown that homeostasis is maintained [7]. One study concluded upon the oscillatory nature of $T_{reg}$ numbers through life, with peaks in adolescence and in over 60 year olds [8]. A recent study in Th17 cells with age showed a small decrease in the frequency of Th17 cells in the CD4+ memory population in elderly donors relative to young [9]. However, for the balance between $T_{reg}$s and Th17 cells, there is currently no published literature.

There is evidence that the balance between $T_{reg}$s and Th 17s is altered in age-related diseases such as acute coronary syndrome [10]. Thus it is intuitive that this balance should be examined with a cross sectional study in healthy donors of different ages. Laboratory experimentation will begin by using flow cytometry to enumerate peripheral blood cells expressing CD4, CD25 and signature transcription factors of these subsets: Foxp3 and Helios for $T_{reg}$s and $ROR_C$ for Th17 cells.

## 3 Method: Simulation of $T_{reg}$ Dynamics

Ultimately we wish to build a model of the dynamics of $T_{reg}$s and Th17 cells throughout life from data currently being collected. Preliminary work for this has involved building a system dynamics simulation in AnyLogic 6.5.0 University Edition, based on a mathematical model by Baltcheva [11]. This model was selected as it comprehensively incorporates the functional dynamics of $T_{reg}$s in terms of homeostasis and during an acute immune response. It characterises the changing precursor and mature $T_{reg}$ populations throughout the human lifetime. Key assumptions include that there is no change in function or responsiveness throughout the lifetime, nor a change to other influential factors on their dynamics such as dendritic cell number and function [11]. Also, the immune response considered includes an expansion and contraction phase, and only one response can occur at a given timepoint [11]. The original model was based on numbers of CD4+CD25±CD45R0− (precursor) and CD4+CD25±CD45R0+ (mature) populations in 119 peripheral blood samples of donors aged 19 to 81 [11]. Although total numbers of CD4+CD25+ cells remained constant, the ratio of precursor to mature was inverted in early adulthood. This represents an important dimension to the observed homeostasis in $T_{reg}$ numbers throughout the lifespan, especially when considering thymic involution from adolescence, reducing the number of new cells entering the system.

Ordinary differential equations describe the dynamics of the above mentioned cells, and stochastic processes control the frequency, duration and antigen-specific nature of primary and secondary immune responses on the different cell compartments [11]. In this work, a simple scenario was chosen in order to test the hypothesis that the model could be implemented in AnyLogic. The scenario assumes a lack of antigen-induced proliferation and death of both precursor and mature $T_{reg}$s, no density-dependent pro-liferation and death and thymic output as the only external input into the various Treg subsets [11]. The parameter values used correspond to means of the distributions for scenario 2iiia given in Baltcheva's work [11]. The simulation is shown below:

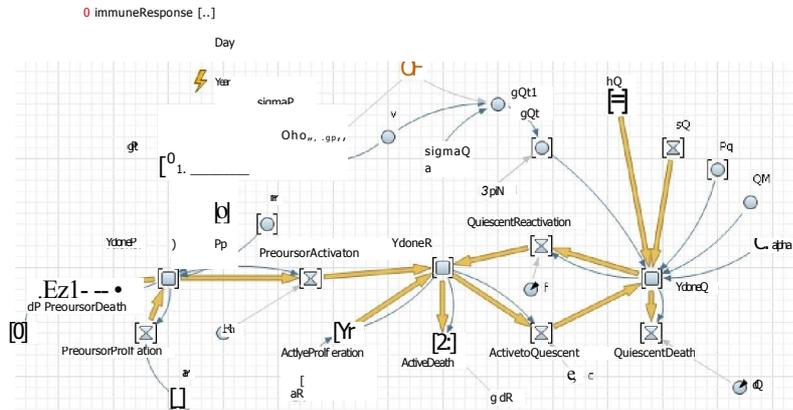

*Figure 1: Main View. Events are shown with for time in days (t) and years (y). Flow variables given by [z] involve a rate of conversion multiplied by the number of cells in the stocks named Y cloneP, —R and —Q. P corresponds to precursors, R to active matures and Q to quiescent matures. These stocks are prefixed by Y clone as they represent total and antigen-specific $T_{reg}$s determined by an array. Homeostatic parameters apply to the total $T_{reg}$ population, whereas immune response parameters are applied to a proportion given by piN. For example, flow of specific precursors into the active mature stock is given by specific precursor stock multiplied by the maturation rate. Each time point, AnyLogic recalculates each stock using the flows defined.*

An additional class (immuneResponse) controls the immune system functional status using *IRstatechart:*

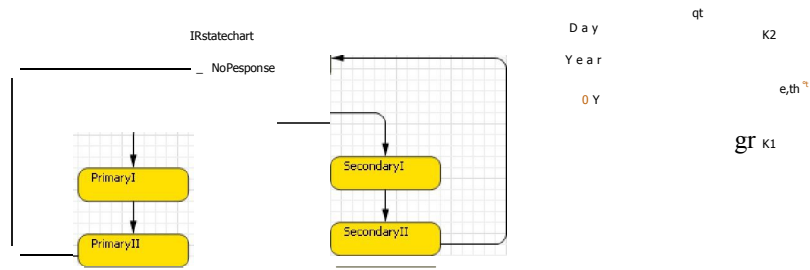

*Figure 2: immuneResponse View. At the beginning of a run the immune system is in the N °Response state. Every 100.95 time steps, the immune system mounts a primary immune response with probability qt or defaults to a secondary response. PrimaryI and SecondaryI, represent the expansion phase. PrimaryI I and SecondaryI I, represent the contraction phase which continues until a new response is instigated. During PrimaryI, parameter b is applied. During PrimaryII, b is set to zero and parame-*

ters c, dR and dQ are applied. In Secondary) parameters b and f are applied and in SecondaryI I these are set to zero and c, dR and dQ are applied.

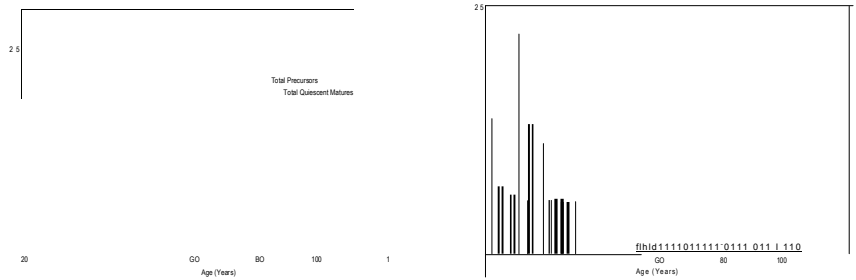

*Figure 3 (generated using MATLAB 122010a): Output data from AnyLogic over a complete run. a) Total precursor and quiescent mature $T_{rey}$ stocks. b) Total active $T_{reg}$s. Each peak corresponds to antigen-specific clones experiencing either primary or secondary immune responses.*

Figure 4 shows simulation output compared to the original dataset. Data has been collected for each stock over 3 complete replications. The maximum standard deviation between three runs for the total precursors was $2.753 \times 10^{-7}$. The maximum standard deviation for total quiescent matures was 9411.

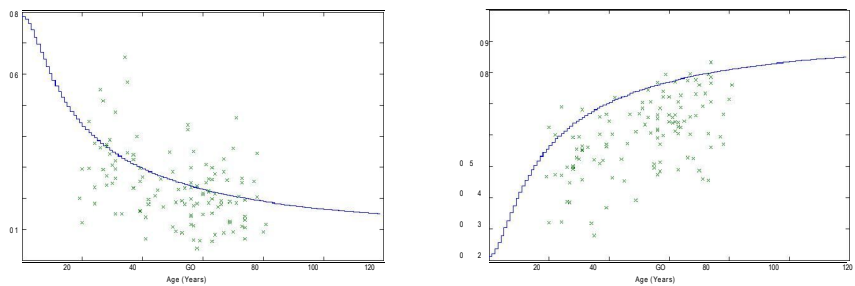

*Figure 4: Output data compared to Baltcheva's dataset. a) The proportion of total $T_{reg}$s in the precursor stock. b) The proportion of total $T_{reg}$s in the quiescent stock.*

In order to quantify how similar the simulation output was to the laboratory data, both datasets were split into 10-year age groups. The median was calculated for these groups, and the difference between medians for simulation output data and laboratory data have been documented below. A Mann Whitney test was then performed for the null hypothesis of no difference between laboratory and output data and the p values are given below:

| Age (Years) | Median Difference | | Mann Whitney Test | |
|---|---|---|---|---|
| | Proportion of Precursors | Proportion of Matures | Proportion of Precursors | Proportion of Matures |
| 10-19 | 0.3101 | 0.0207 | p=0.083 | p=0.658 |
| 20-29 | 0.0525 | 0.1092 | p<0.001 | p<0.001 |
| 30-39 | 0.0290 | 0.1227 | p=0.034 | p<0.001 |
| 40-49 | 0.0690 | 0.1182 | p=0.001 | p<0.001 |
| 50-59 | 0.0587 | 0.1007 | p<0.001 | p<0.001 |
| 60-69 | 0.0072 | 0.1230 | p=0.127 | p<0.001 |
| 70-79 | 0.0011 | 0.1204 | p=0.808 | p<0.001 |
| 80-89 | 0.0805 | 0.0775 | p=0.014 | p=0.014 |

*Table 1: Comparison of median output and laboratory data for each 10-year age group.*

## 4 Discussion and Concluding Remarks

The implementation of Baltcheva's model as a system dynamics simulation has been documented here and compared to experimental evidence. It has been shown that the simulation mimics the key feature of inversion of precursor and memory cells in early adulthood. The lack of statistical similarity between simulation output and laboratory data indicates that further validation of this model is necessary and will involve a comparison of other scenarios proposed in Baltcheva's work. Ultimately we will develop and validate a simulation of our novel dataset of $T_{reg}$ and Th17 cells using this sort of approach. CD4+CD25±Foxp3+ and CD4±Foxp3±Helios+ cell numbers instead of CD4+CD25+ s for $T_{reg}$ s will also be collected as they are arguably more specific markers [12][13]. In terms of improving the simulation, alternatives to continually reactivating a single $T_{reg}$ clone are required, as is simulation of more than one immune stimulus at a time. Baltcheva discloses various assumptions including no difference in $T_{reg}$ function [11]. It may be possible to improve this model by considering functional as well as numerical changes to $T_{reg}$ subsets with age.

A more abstract research question is whether a simplistic model of immunosenescence can lend useful insight into the biological problem. It can be argued that the process of simulation alone might allow researchers to address assumptions and allow for systematic generation of hypotheses. Also, hypotheses which are difficult to test in the laboratory might be testable with a simulation. For example, we might introduce an intervention to mimic ablative chemotherapy by depleting each stock at a single time point. Total values of each stock might then be compared for simulation runs with or without intervention to make hypotheses about $T_{reg}$ recovery. Simulating the dynamics of Th17 cells in parallel to $T_{reg}$ s may also allow us to make predictions about the maintenance of their balance throughout life, would allow for extreme parameter values to be tested and may indicate a maximum length of time for homeostasis to be maintained.

Our primary hypothesis is that age alters $T_{reg}$ and Th17 cells with consequences for health in older age and we aim to conduct a cross sectional study to obtain the distribution of particular changes. We anticipate that a strategy of both laboratory investigation and system dynamics simulation as exemplified by Baltcheva's work will be useful to address relationships between T cell subsets over time. The model might also be developed to consider new questions about response to interventions and the length of time the immune system might be able to maintain $T_{reg}$ and Th17 cell homeostasis.

*With thanks to Irina Baltcheva for providing her raw data.*